\documentclass[12pt]{iopart}
\usepackage{epsfig}
\usepackage{iopams}
\usepackage{color}
\usepackage{enumitem}

\newcommand{\be}{\begin{equation}}
\newcommand{\ee}{\end{equation}}

\newcommand{\ba}{\begin{array}}
\newcommand{\ea}{\end{array}}
\newcommand{\bea}{\begin{eqnarray}}
\newcommand{\eea}{\end{eqnarray}}

\newtheorem{corollary}{Corollary}
\newtheorem{proposition}{Proposition}
\newtheorem{lemma}{Lemma}

\begin{document}

\title[Minimal surfaces in the soliton surface approach]{Minimal surfaces in the soliton surface approach}
\author{A Doliwa$^1$ and A M Grundland$^{2,3}$}

\address{$^1$ Faculty of Mathematics and Computer Science, University of Warmia and Mazury, S\l oneczna 54, 10-710 Olsztyn, Poland}
\address{$^2$ Centre de Recherches Math\'ematiques, Universit\'e de Montr\'eal,\\ Montr\'eal CP 6128 (QC) H3C 3J7, Canada}
\address{$^3$ Department of Mathematics and Computer Science, Universit\'e du Qu\'ebec, Trois-Rivi\`eres, CP 500 (QC) G9A 5H7, Canada}
\ead{doliwa@matman.uwm.edu.pl, grundlan@crm.umontreal.ca}

\begin{abstract}
The main objective of this paper is to derive the Enneper-Weierstrass representation of minimal surfaces in $\mathbb{E}^3$ using the soliton surface approach. We exploit the Bryant-type representation of conformally parametrized surfaces in the hyperbolic space $H^3(\lambda)$ of curvature $-\lambda^2$, which can be interpreted as a 2 by 2 linear problem involving the spectral parameter $\lambda$. In the particular case of constant mean curvature-$\lambda$ surfaces a special limiting procedure $(\lambda\rightarrow0)$, different from that of Umehara and Yamada \cite{UY}, allows us to recover the Enneper-Weierstrass representation. Applying such a limiting procedure to the previously known cases, we obtain Sym-type formulas. Finally we exploit the relation between the Bryant representation of constant mean curvature-$\lambda$ surfaces and second-order linear ordinary differential equations. We illustrate this approach by the example of the error function equation.

\paragraph{}Keywords: Integrable systems, minimal surfaces, Weierstrass representation, soliton surfaces. 
\end{abstract}

\maketitle

\section{Introduction}\setcounter{equation}{0}
The theory of minimal surfaces in Euclidean three-space $\mathbb{E}^3$ goes back to Joseph-Louis Lagrange in 1768, and is a classical subject of differential geometry (see e.g. \cite{Bianchi,15}). We recall that, given a variation of the surface $S$ along the vector field $\vec{\nu}$ vanishing on its boundary, the corresponding variation of the area of $S$, up to higher-order terms in the small parameter $\epsilon$, is given by
\be
A(S+\epsilon\vec{\nu})-A(S)=-2\epsilon\int_S\vec{\nu}\cdot \vec{H}dA+...,
\ee 
where $\vec{H}$ is the mean curvature vector on the surface. Therefore surfaces with vanishing mean curvature are called minimal surfaces. It turns out that, due to the Enneper-Weierstrass formula, one can construct minimal surfaces in terms of two meromorphic functions \cite{E,W,Spirak}.

\subsection{Soliton surfaces}
In the last three decades, many special classes of surfaces have been studied using methods of soliton theory \cite{Bob94,Kono,19}. In the work of Bobenko \cite{Bob94} one can find a list of such integrable surfaces.
Integrable equations in two independent variables result from the compatibility condition (the Zakharov-Shabat equations \cite{ZS})
\be
U_{,y}(\lambda)-V_{,x}(\lambda)+[U(\lambda),V(\lambda)]=0,\label{ZSeq}
\ee
of two linear equations (a spectral problem)
\be
\Psi_{,x}=U(\lambda)\Psi,\qquad \Psi_{,y}=V(\lambda)\Psi,
\ee
where, in this context, $\lambda$ is called the spectral parameter. When a solution $\Psi(x,y;\lambda)$ takes values in a Lie group $G$ and $U(x,y;\lambda)$, $V(x,y;\lambda)$ are matrix functions in the associated Lie algebra $\mathfrak{g}$, then the immersion function
\be
F(x,y;\lambda)=\Psi^{-1}(x,y;\lambda)\Psi(x,y;\lambda)_{,\lambda}\label{1.1}
\ee
of the variables $x$, $y$ can be interpreted, for a fixed $\lambda\in\mathbb{C}$, as a surface in the Lie algebra $\mathfrak{g}$, provided that the tangent vectors
\be
F_{,x}=\Psi^{-1}U_{,\lambda}\Psi,\qquad F_{,y}=\Psi^{-1}V_{,\lambda}\Psi
\ee
are linearly independent. Such a formula, which first appeared in the works of A Sym \cite{18,19} for surfaces immersed in semisimple Lie algebras \cite{10}-\cite{12} and subsequently was used by A Bobenko in \cite{3}-\cite{bo4}, allows the establishment of a backward link between geometry and integrable systems. 
For integrable equations coming from surface theory, the spectral parameter $\lambda$ describes deformations within a certain class of surfaces, that is, integrable surfaces always appear as one-parameter families of surfaces \cite{17}.

Since then the applicability of the Sym formula (sometimes called the Sym-Tafel immersion formula, see footnote $(^{19})$ in \cite{18}), to geometric problems related to soliton equations has been extended. In particular, new terms have been added to its original form corresponding to a gauge symmetry of the linear spectral problem (LSP) \cite{4,5,DS92} and generalized symmetries of the zero-curvature representation of integrable nonlinear PDEs \cite{12,13,14}. 
Finding a list among such immersions of those which have an invariant geometric characterization would be important both for a geometric interpretation of surfaces as well as for various applications defined by some restrictions on the arbitrary functions. For instance it was shown that the Fokas-Gel'fand approach \cite{12,13} can be expressed in the framework of the Sym approach and the two approaches are equivalent \cite{4}. The latter formula is also a suitable tool to determine and construct discrete surfaces on the lattice \cite{bo4,Bob}.

In addition to many fruitful applications of the soliton-surface approach there are still some classical classes of surfaces, including the minimal surfaces, which have not been incorporated into the scheme. We remark that, apart from the Lie algebra interpretation of the Sym formula, there exists another interpretation related to Clifford algebras, where the spectral parameter $\lambda$ is considered in relation to the curvature of the ambient space \cite{6,9}. The differentiation with respect to the spectral parameter $\lambda$ then appears as a result of a limiting procedure and L'H\^opital's rule.

In the present work we incorporate the theory of minimal surfaces into the soliton surface theory. We start, in section 2, from the spinorial interpretation of conformally parametrized surfaces in the hyperbolic space $H^3(\lambda)$ of curvature $-\lambda^2$. In section 3 the Clifford algebra interpretation in the limiting procedure recovers the classical Enneper-Weierstrass formula in the case of constant mean curvature (CMC)-$\lambda$ surfaces. In section 4 we discuss an interesting link between the Bryant CMC-$\lambda$ surfaces \cite{Br,Br2} and a certain second-order linear ordinary differential equation (ODE). The latter connection is illustrated by an example of CMC-$\lambda$ surfaces constructed from a solution of the error function equation.

\subsection{Minimal surfaces and their Weierstrass representation}
Let $F:\mathcal{R}\rightarrow\mathbb{E}^3$ be a conformal minimal immersion of a Riemann surface $\mathcal{R}$. Then the one-forms $\varphi_k=\partial F_k$, $k=1,2,3$ are holomorphic, have no real periods and satisfy the equations $\sum_{k=1}^3\varphi_k^2=0$. The intrinsic metric in $\mathcal{R}$ is given by $ds^2=\sum_{k=1}^3\vert\varphi_k\vert^2$, hence the one forms $\varphi_k$'s have no common zeroes \cite{Yang}.

Conversely, any vectorial holomorphic one-form $\varphi=(\varphi_1,\varphi_2,\varphi_3)$ on $\mathcal{R}$ without real periods and satisfying $\sum_{k=1}^3\varphi_k^2=0$ and $\sum_{k=1}^3\vert\varphi_k(P)\vert^2\neq0$ for all points $P\in\mathcal{R}$, determines a conformal immersion $F:\mathcal{R}\rightarrow\mathbb{E}^3$ by the expression
\be
F=\mbox{Re}\left(\int_{z_0}^z\varphi\right).\label{1.2}
\ee
The meromorphic function $\psi=\varphi_3(\varphi_1-i\varphi_2)^{-1}$ corresponds to the Gauss map of a smooth orientable surface $F$ in $\mathbb{E}^3$ up to the stereographic projection, and $\varphi$ can be written as \cite{16}
\be
\varphi=\left(\frac{1}{2}(1/\psi-\psi),\frac{i}{2}(1/\psi+\psi),1\right)\varphi_3.\label{1.3}
\ee

\textbf{Remark.} In the local representation of $\varphi_3=\psi\eta^2dz$, where $\eta$ is a local holomorphic function in $\mathbb{E}^3$ of the complex variable $z\in\mathbb{C}$, we obtain the standard form of the Enneper-Weierstrass representation of minimal surfaces
\be
F=\mbox{Re}\left[\int_{z_0}^z\left(\frac{1}{2}(1-\psi^2),\frac{i}{2}(1+\psi^2),\psi\right)\eta^2dz'\right].\label{1.4}
\ee
The three components of $F(z,\bar{z})$ can be identified as the coordinates of a minimal surface in $\mathbb{E}^3$ \cite{K}. 

\section{Conformally immersed surfaces in hyperbolic three-space}\setcounter{equation}{0}
To describe minimal surfaces in the soliton surfaces approach we need to consider a closely related special class of CMC surfaces in three-dimensional hyperbolic space $H^3$. We start therefore by recalling the standard description of conformally parametrized generic surfaces in $H^3$ (see e.g. \cite{Bob}).

\subsection{Conformal immersions of surfaces in $H^3$ and the corresponding equations}
The formulas presented in this Section are fairly standard from the geometric point of view. In the context of the application of methods of soliton theory to the study of CMC surfaces in the hyperbolic space $H^3(\lambda)$, the spinor representation of the GW equations was used by Bobenko in \cite{3}.
Consider the Lorentz space $\mathbb{R}^{3,1}$ with the standard bilinear form
\be
(X\vert Y)=X_1Y_1+X_2Y_2+X_3Y_3-X_0Y_0.
\ee
We denote by $H^3(\lambda)\subset\mathbb{R}^{3,1}$ the three-dimensional hyperboloid given by the equations $(X\vert X)=-\lambda^{-2}$. On $H^3(\lambda)$ the induced metric is positive definite and has constant sectional curvature.

Given a conformal immersion $F:\mathcal{R}\rightarrow H^3(\lambda)\subset\mathbb{R}^{3,1}$ of the Riemann surface $\mathcal{R}$, with the local complex coordinate $z=x+iy$, we have
\be
(F_{,z}\vert F_{,z})=(F_{,\bar{z}}\vert F_{,\bar{z}})=0,
\ee
and also
\be
(F_{,z}\vert F)=(F_{,\bar{z}},F)=0.
\ee
Supplement the vectors $F$, $F_{,z}$, $F_{,\bar{z}}$ with the unit normal $N$
\be
(F\vert N)=(F_{,z}\vert N)=(F_{,\bar{z}}\vert N)=0,\qquad (N\vert N)=1,
\ee
and define the functions $u$, $H$ and $Q$ by
\be
(F_{,z}\vert F_{,\bar{z}})=\frac{1}{2}e^u,\qquad (F_{,z\bar{z}}\vert N)=\frac{1}{2}He^u,\qquad (F_{,zz}\vert N)=Q.
\ee
Then the non-trivial part of the Gauss-Weingarten (GW) equations of the moving frame takes the form
\bea
F_{,zz}=u_{,z}F_{,z}+QN,\label{2.1}\\
F_{,z\bar{z}}=\frac{\lambda^2}{2}e^uF+\frac{1}{2}He^uN,\label{2.2}\\
N_{,z}=-HF_{,z}-2Qe^{-u}F_{,\bar{z}}.\label{2.3}
\eea
The GMC equations then read
\be
u_{,z\bar{z}}+\frac{1}{2}(H^2-\lambda^2)e^u-2\vert Q\vert^2e^{-u}=0,\qquad Q_{,\bar{z}}=\frac{1}{2}H_{,z}e^u.\label{2.4}
\ee

\subsection{Spinor formalism}
Identify the Lorentz space with the $2\times2$ Hermitian matrices
\be\hspace{-2.5cm}
X=(X_0,X_1,X_2,X_3)\leftrightarrow X^\sigma=X_0\mathbb{I}_2+\sum_{k=1}^3X_k\sigma_k=\left(\ba{cc}
X_0+X_3 & X_1-iX_2 \\
X_1+iX_2 & X_0-X_3
\ea\right),\label{2.5}
\ee
where $\mathbb{I}_2$ is the $2\times2$ identity matrix and $\sigma_k$ are the Pauli matrices
\be
\sigma_1=\left(\ba{cc}
0 & 1 \\
1 & 0
\ea\right),\qquad\sigma_2=\left(\ba{cc}
0 & -i \\
i & 0
\ea\right),\qquad\sigma_3=\left(\ba{cc}
1 & 0 \\
0 & -1
\ea\right).
\ee
The scalar product of vectors in terms of matrices is given by
\be\hspace{-2.5cm}
(X\vert Y)=\frac{1}{2}\tr(X^\sigma\epsilon(Y^\sigma)^T\epsilon),\qquad \epsilon=\left(\ba{cc}
0 & 1 \\
-1 & 0
\ea\right)=i\sigma_2,\qquad (X\vert X)=-\det X^\sigma.\label{inner}
\ee
In constructing the $2\times2$ matrix representation of the GW equations (\ref{2.1})-(\ref{2.3}) we use the homomorphism $\rho:SL(2,\mathbb{C})\rightarrow SO(3,1)$ given by
\be
(\rho(a)X)^\sigma=a^\dagger X^\sigma a,\label{rhoa}
\ee
where $a^\dagger$ denotes the hermitian conjugate of $a$. In the reduction from the Lorentz space to the Euclidean space $\mathbb{E}^3$, obtained by putting $X_0=0$, the corresponding rotation is given by an element $a\in SU(2)$ of the special unitary group.

In order to write the GW equations (\ref{2.1})-(\ref{2.3}) in the spinor formalism we look for a $SL(2,\mathbb{C})$-valued function $\Phi$ which transforms the orthonormal (with respect to the scalar product (\ref{inner})) basis $(\mathbb{I}_2,\sigma_1,\sigma_2,\sigma_3)$ into the orthonormal basis
\be
\left(\lambda F^\sigma,e^{-u/2}F^\sigma_{,x},e^{-u/2}F^\sigma_{,y},N^\sigma\right)=\Phi^\dagger(\mathbb{I}_2,\sigma_1,\sigma_2,\sigma_3)\Phi,\label{2a1}
\ee
where $F^\sigma$ is defined according to (\ref{2.5}). Then we have
\be
F^\sigma_{,z}=e^{u/2}\Phi^\dagger\left(\ba{cc}
0 & 0 \\
1 & 0
\ea\right)\Phi,\qquad F^\sigma_{,\bar{z}}=e^{u/2}\Phi^\dagger\left(\ba{cc}
0 & 1 \\
0 & 0
\ea\right)\Phi.\label{2a2}
\ee

If we define the $\mathfrak{sl}(2,\mathbb{C})$-valued functions $U$, $V$ by
\be
\Phi_{,z}=U\Phi,\qquad \Phi_{,\bar{z}}=V^\dagger\Phi,\label{2.6}
\ee
then we also have
\be
\Phi^\dagger_{,z}=\Phi^\dagger V,\qquad \Phi^\dagger_{,\bar{z}}=\Phi^\dagger U^\dagger.
\ee

\begin{lemma}
Using the homomorphism (\ref{rhoa}), the moving frame $(F,F_{,z},F_{,\bar{z}},N)^T$ of the conformally parametrized surface is described by the formulae (\ref{2.1})-(\ref{2.3}) where the wavefunction $\Phi\in SL(2,\mathbb{C})$ satisfies the equations (\ref{2.6}) and the $\mathfrak{sl}(2,\mathbb{C})$-valued functions $U$ and $V$ are of the form
\be
\hspace{-1.5cm}U=\left(\ba{cc}
\frac{1}{4}u_{,z} & -Qe^{-u/2} \\
\frac{1}{2}e^{u/2}(\lambda+H) & -\frac{1}{4}u_{,z}
\ea\right),\qquad V=\left(\ba{cc}
-\frac{1}{4}u_{,z} & Qe^{-u/2} \\
\frac{1}{2}e^{u/2}(\lambda-H) & \frac{1}{4}u_{,z}
\ea\right).\label{UV}
\ee
\end{lemma}

\textit{Proof.} Making use of (\ref{2a1}) and (\ref{2a2}) we can express the derivative $F^\sigma_{,z}$ (the trivial part of the GW equations) in two ways
\be
F^\sigma_{,z}=\frac{1}{\lambda}\Phi^\dagger(V+U)\Phi=e^{u/2}\Phi^\dagger\left(\ba{cc}
0 & 0 \\
1 & 0
\ea\right)\Phi
\ee
to obtain
\be
V_{11}+U_{11}=V_{12}+U_{12}=V_{22}+U_{22}=0,\qquad V_{21}+U_{21}=\lambda e^{u/2}.
\ee
Similarly, making use of the other GW equations, we derive the final form of the matrices.

\hfill $\square$

\begin{corollary}
The zero curvature representation for (\ref{2.7})
\be
U_{,\bar{z}}-V_{,z}^\dagger+[U,V^\dagger]=0,
\ee
is equivalent to the GMC equations.
\end{corollary}

\subsection{The Sym-type immersion formula in $H^3(\lambda)$}
We emphasize the following fact which is a direct consequence of the representation (\ref{2a1})
\begin{proposition}
Given a solution $(u,Q,H)$ of the GMC equations (\ref{2.4}), and given an $SL(2,\mathbb{C})$-valued solution $\Phi$ of the linear system (\ref{2.6}) with matrices as in Lemma~1, the immersion function
\be
F^\sigma=\frac{1}{\lambda}\Phi^\dagger\Phi,\label{2.7}
\ee
represents a conformal immersion in $H^3(\lambda)$.
\end{proposition}

Formula (\ref{2.7}) plays an essential role in deriving the Weierstrass-Enneper formula for minimal surfaces in $\mathbb{E}^3$ and will be used in what follows. Notice that in the limit $\lambda\rightarrow0$, the hyperbolic space $H^3$ becomes the standard Euclidean space $\mathbb{E}^3$. 
However we cannot take the direct limit $\lambda\rightarrow0$ at the level of the representation formula (\ref{2.7}). Following \cite{10}, before taking the limit, we first shift the origin from the center of the hyperboloid to one of its points, which does not change the geometry of the immersion under consideration. This results in the ``finite'' formula (i.e. this does not involve infinity in the immersion formula $\tilde{F}^\sigma$)
\be
\tilde{F}^\sigma=\lim_{\lambda\rightarrow0}\frac{1}{\lambda}(\Phi^\dagger\Phi-\mathbb{I}_2).\label{2.8}
\ee
An analogous procedure allows us to recover the Sym formula in the case of integrable kinematics of curves \cite{10,GG} and the corresponding Ablowitz-Ladik equation \cite{AL}, where a similar application of the Clifford algebra and spinorial representation of the orthogonal group gives, after the application of L'H\^opital's rule, an explanation of the differentiation with respect to the spectral parameter $\lambda$ (see also subsequent works \cite{5,6} on analogous results for $n$-dimensional spaces of constant negative curvature in $\mathbb{R}^{2n-1}$ and CMC surfaces).

\section{Special CMC-surfaces in $H^3$ and the Weierstrass representation}\setcounter{equation}{0}
\subsection{$H=\lambda$ surfaces in $H^3(\lambda)$}
From now on we will only be interested in special CMC surfaces where $H=\lambda$. The reason is that in such cases the GMC equations reduce to the same system
\be
u_{,z\bar{z}}-2\vert Q\vert^2e^{-u}=0,\qquad Q_{,\bar{z}}=0,\label{3.1}
\ee
as in the case of minimal surfaces in $\mathbb{E}^3$. Notice that such a situation is not possible in the case of conformal immersions in the sphere $S^3$ of radius $1/\lambda$ embedded in $\mathbb{E}^4$. 
The formal transition between $H^3(\lambda)$ and $S^3(\lambda)$ can be obtained by replacing $\lambda$ by $i\lambda$ in the GW and the GMC equations (\ref{2.1})-(\ref{2.3}).

In looking for solutions of the reduced linear problem for $\Phi\in SL(2,\mathbb{C})$
\be
\Phi_{,z}=\left(\ba{cc}
\frac{1}{4}u_{,z} & -Qe^{-u/2} \\
\lambda e^{u/2} & -\frac{1}{4}u_{,z}
\ea\right)\Phi,\qquad\Phi_{,\bar{z}}=\left(\ba{cc}
-\frac{1}{4}u_{,\bar{z}} & 0 \\
\bar{Q}e^{-u/2} & \frac{1}{4}u_{,\bar{z}}
\ea\right)\Phi,\label{3.2}
\ee
we will be guided by the corresponding facts from the theory of minimal surfaces in $\mathbb{E}^3$.

The following result is implied by the Weierstrass representation of minimal surfaces (\ref{1.4}) together with (\ref{2.5}).
\begin{lemma}
The general solution of the reduced system (\ref{3.1}) in terms of two arbitrary holomorphic functions $\eta$, $\psi$ has the form
\be
e^{u/2}=\eta\bar{\eta}(1+\psi\bar{\psi}),\qquad Q=-\eta^2\psi_{,z}.\label{3.3}
\ee
\end{lemma}
When $Q\equiv1$ the GMC equation (\ref{3.1}) reduces to the Liouville equation, which has the well-known solution
\be
e^u=\vert\psi_{,z}\vert^{-2}(1+\vert\psi\vert^2)^2.
\ee

It is also convenient to simplify the reduced linear problem (\ref{3.2}) by following the gauge transform $\Psi=M\Phi$, where
\be
M=\frac{1}{(1+\psi\bar{\psi})^{1/2}}\left(\ba{cc}
\left(\frac{\eta}{\bar{\eta}}\right)^{1/2}\psi & -\left(\frac{\eta}{\bar{\eta}}\right)^{1/2} \\
\left(\frac{\bar{\eta}}{\eta}\right)^{1/2} & \left(\frac{\bar{\eta}}{\eta}\right)^{1/2}\bar{\psi}
\ea\right)\in SU(2).\label{3.4}
\ee
\begin{lemma}
After applying the gauge transformation $M$ given by (\ref{3.4}) to the wavefunction $\Phi$, we obtain that the function $\Psi=M\Phi$ satisfies the following linear system
\be
\Psi_{,z}=\lambda\eta^2\left(\ba{cc}
\psi & -1 \\
\psi^2 & -\psi
\ea\right)\Psi,\qquad \Psi_{,\bar{z}}=0.\label{3.5}
\ee
\end{lemma}

\textit{Proof.} The result can be verified by direct calculation using the standard formulas
\be
\ba{l}
\Psi_{,z}=(M_{,z}M^{-1}+MUM^{-1})\Psi,\\
\Psi_{,\bar{z}}=(M_{,\bar{z}}M^{-1}+MV^+M^{-1})\Psi.
\ea\label{LSPG}
\ee 
It is however instructive to decompose the gauge matrix $M$ into a sequence of transformations
\be
M=\left(\ba{cc}
1 & 1/\psi \\
0 & 1
\ea\right)\left(\ba{cc}
\eta\psi & 0 \\
0 & 1/(\eta\psi)
\ea\right)\left(\ba{cc}
1 & 0 \\
\alpha^{-1} & 1
\ea\right)\left(\ba{cc}
e^{u/4} & 0 \\
0 & e^{-u/4}
\ea\right),
\ee
where $\alpha=\eta^2\psi(1+\psi\bar{\psi})$, and simplify the linear system step by step.\hfill$\square$

\textbf{Remark.} The gauge matrix $M$ is equal to the inverse of the fundamental solution of the linear system (\ref{3.2}) for $\lambda=0$. The latter corresponds to minimal surfaces in $\mathbb{E}^3$ and provides an explanation for the fact that $M$ takes values in the $Spin(3)=SU(2)$ group.

Recall \cite{Maurin} that by the method of successive approximations, the formal solution of the ordinary linear matrix equation satisfying the initial-value problem
\be
\Theta'(t)=H(t)\Theta(t),\qquad \Theta(t_0)=\mathbb{I},
\ee
is given in the form of a formal series
\be
\ba{l}
\Theta(t)=\mathbb{I}+\int_{t_0}^tdt_1H(t_1)+\int_{t_0}^tdt_2\int_{t_0}^{t_2}dt_1H(t_2)H(t_1)+...\\
+\int_{t_0}^tdt_k\int_{t_0}^{t_k}dt_{k-1}...\int_{t_0}^{t_2}dt_1H(t_k)H(t_{k-1})...H(t_1)+...
\ea
\ee
Using the above formula and the decomposition of the rank $1$ holomorphic matrix
\be
\left(\ba{cc}
\psi & -1 \\
\psi^2 & -\psi
\ea\right)=\left(\ba{c}
1\\
\psi
\ea\right)(\ba{cc}
\psi, & -1
\ea),\qquad \psi_{,\bar{z}}=0\label{3.10a}
\ee
we obtain the formal solution of the linear system.

\begin{lemma}
The formal fundamental solution of the reduced linear system (\ref{3.5}), obtained by successive iterations by the gauge transformation, is given in the form of a series
\be\hspace{-2.5cm}
\ba{l}
\Psi(z)=\mathbb{I}_2+\lambda\int_{z_0}^zdz_1\eta(z_1)^2\left(\ba{cc}
\psi(z_1) & -1 \\
\psi(z_1)^2 & -\psi(z_1)
\ea\right)+...\\
+\lambda^k\int_{z_0}^zdz_k\int_{z_0}^{z_2}dz_1\prod_{i=1}^k\eta_{z_i}^2\prod_{i=1}^{k-1}(\psi(z_{i+1}-\psi(z_i))\left(\ba{cc}
\psi(z_1) & -1 \\
\psi(z_1)\psi_{z_k} & -\psi(z_k)
\ea\right)+...
\ea\label{3a1}
\ee
\end{lemma}

\subsection{The limit $\lambda\rightarrow0$ applied to minimal surfaces in $\mathbb{E}^3$}
To obtain the limiting case of minimal surfaces in $\mathbb{E}^3$ we apply formula (\ref{2.8}) to a solution $\Phi$ of the linear problem (\ref{3.2}) because of the decomposition $\Psi=M\Phi$ discussed in Lemma 3. Equation (\ref{2.8}) provides a deformation of the Bryant representation \cite{Br,Br2} of $H=\lambda$ surfaces in $H^3(\lambda)$ to the Weierstrass formula, which is different from the deformation given by Umehara and Yamada \cite{UY}. Their approach used the intermediate-step stereographic projection.

For our purpose it is therefore enough to consider the first two terms $\Psi=\mathbb{I}_2+\lambda\Psi_1+...$ in the series (\ref{3a1}) of Lemma~4. 
In the limit we obtain the following Clifford algebra representation of the minimal immersion
\be
\hspace{-2.5cm}\tilde{F}^\sigma=\Psi_1+\Psi_1^+=\left(\ba{cc}
\int^z\eta^2\psi d\zeta+\overline{\int^z\eta^2\psi d\zeta} & -\int^z\eta^2d\zeta+\overline{\int^z\eta^2\psi^2d\zeta} \\
\int^z\eta^2\psi^2d\zeta-\overline{\int^z\eta^2d\zeta} & -\int^z\eta^2\psi d\zeta-\overline{\int^z\eta^2\psi d\zeta}
\ea\right)\in\mathfrak{sl}(2,\mathbb{C}),\label{3.6}
\ee

Using formula (\ref{2.5}), equation (\ref{3.6}) gives the Cartesian coordinates of the corresponding minimal surfaces in $\mathbb{E}^3$.
It constitutes the standard form (up to minor changes) of the Weierstrass-Enneper representation of minimal surfaces in $\mathbb{E}^3$
\bea
F_1=\frac{1}{2}Re\left(\int^z(\psi^2-1)\eta^2d\zeta\right),\\
F_2=\frac{1}{2i}Im\left(\int^z(\psi^2+1)\eta^2d\zeta\right),\\
F_3=2Re\left(\int^z\psi\eta^2d\zeta\right).
\eea

\section{Examples of second-order linear differential equations arising from the CMC-$\lambda$ surfaces}\setcounter{equation}{0}
In this section we discuss a closed connection of equation (\ref{3.5}) with a second-order ordinary scalar linear differential equation in the complex domain.

Consider a one-column version of equation (\ref{3.10a})
\be
\partial\left(\ba{c}
\alpha\\
\beta
\ea\right)=\lambda\eta^2\left(\ba{cc}
\psi & -1 \\
\psi^2 & -\psi
\ea\right)\left(\ba{c}
\alpha\\
\beta
\ea\right),\qquad \bar{\partial}\left(\ba{c}
\alpha\\
\beta
\ea\right)=\left(\ba{c}
0\\
0
\ea\right),\label{4.1}
\ee
where $\partial=\partial/\partial z$ and $\bar{\partial}=\partial/\partial\bar{z}$, $\eta,\psi$ are meromorphic functions and $\lambda\in\mathbb{C}$. We denote by $\alpha_0,\beta_0$ the initial data for $z=z_0$. Equation (\ref{4.1}) can be reduced to the second-order ODE
\be
\frac{d^2\alpha}{dz^2}-2\frac{(\partial\eta)}{\eta}\frac{d\alpha}{dz}-\lambda\eta^2(\partial\psi)\alpha=0,\label{5.6}
\ee
where $\beta$ satisfies the equation
\be
\beta=\psi\alpha-\frac{1}{\lambda\eta^2}\partial\alpha.
\ee
The differential equation (\ref{5.6}) can also be transformed into its standard form
\be
\frac{d^2y}{dz^2}+Q(z,\lambda)y=0,\qquad Q(z,\lambda)=\partial^2(\ln\eta)-(\partial\ln\eta)^2-\lambda\eta^2(\partial\psi)\label{LP}
\ee
after the change of variable $y=\eta_0\frac{\alpha}{\eta},$ $\eta_0\in\mathbb{C}$.

Let us now perform a reformulation of a selected second-order linear differential equation (\ref{5.6}) associated with the error function equation \cite{Abramowitz}

\be
\frac{d^2w}{dz^2}-2z\frac{dw}{dz}-2nw=0,\qquad n\in\mathbb{Z}
\ee
and comparing its coefficients with the coefficients of the differential equation (\ref{5.6}) and integrating them, we get
\be
\eta=ce^{z^2/2},\qquad Q=1+2n-z^2,\qquad c\in\mathbb{C}.\nonumber
\ee
where the function $\psi$ is given by the error function
\be
\psi=\frac{n\sqrt{\pi}}{\lambda c^2}\mbox{Erf}(z)-c_1,\qquad c_1\in\mathbb{C}.\nonumber
\ee
Hence the potential matrix $U$ given by (\ref{3.10a}), becomes
\be
U=\left(\ba{cc}
\frac{n}{\lambda c^2}\sqrt{\pi}\mbox{Erf}(z) & -1 \\
\frac{n^2}{\lambda^2c^4}\pi\left(\mbox{Erf}(z)\right)^2 & -\frac{n}{\lambda c^2}\sqrt{\pi}\mbox{Erf}(z)
\ea\right).\label{5.11}
\ee
The solution of the LSP (\ref{LSPG}) with the potential matrix given by (\ref{5.11}), can be expressed in terms of the Hermite polynomial $H_{-n}(z)$ and the Kummer confluent hypergeometric function $\hspace{2mm}F_{\hspace{-4mm}1\phantom{ff}1}(\frac{n}{2},\frac{1}{2},z^2)$
\be
\Psi=\left(\ba{cc}
\alpha_1 & \alpha_2 \\
\beta_1 & \beta_2
\ea\right),
\ee
where we have introduced the following notation
\be
\hspace{-2cm}\ba{l}
\alpha_1=e^{-z^2}(H_{-n-1}(z)+\sigma\hspace{2mm}F_{\hspace{-4mm}1\phantom{ff}1}(\frac{1+n}{2},\frac{1}{2},z^2)),\\
\alpha_2=e^{-z^2}(H_{-n-1}(z)+\hspace{2mm}F_{\hspace{-4mm}1\phantom{ff}1}(\frac{1+n}{2},\frac{1}{2},z^2)),\\
\beta_1=\frac{1}{\lambda c^2}e^{-z^2}\left[(\lambda c^2c_1+n\sqrt{\pi}\mbox{Erf}(z))(H_{-n-1}(z)+\sigma\hspace{2mm}F_{\hspace{-4mm}1\phantom{ff}1}(\frac{1+n}{2},\frac{1}{2},z^2))\right]\\
\hspace{1cm}+2e^{z^2}\left[(1+n)H_{-n-2}(z)+zH_{-n-1}(z)-nz\sigma\hspace{2mm}F_{\hspace{-4mm}1\phantom{ff}1}(\frac{1+n}{2},\frac{3}{2},z^2)\right],\\
\beta_2=\frac{1}{\lambda c^2}e^{-z^2}\left[(\lambda c^2c_1+n\sqrt{\pi}\mbox{Erf}(z))(H_{-n-1}(z)+\hspace{2mm}F_{\hspace{-4mm}1\phantom{ff}1}(\frac{1+n}{2},\frac{1}{2},z^2))\right]\\
\hspace{1cm}+2e^{z^2}\left[(1+n)H_{-n-2}(z)+zH_{-n-1}(z)-nz\hspace{2mm}F_{\hspace{-4mm}1\phantom{ff}1}(\frac{1+n}{2},\frac{3}{2},z^2)\right],\\
\sigma=1+\lambda e\left[2((1+n)H_{-n-2}(1)+H_{-n-1}(1))\hspace{2mm}F_{\hspace{-4mm}1\phantom{ff}1}(\frac{1+n}{2},\frac{1}{2},1)\right.\\
\hspace{2cm}\left.+2nH_{-n-1}(1)\hspace{2mm}F_{\hspace{-4mm}1\phantom{ff}1}(\frac{1+n}{2},\frac{3}{2},1)\right]^{-1}
\ea\label{4.c}
\ee
The immersion function $F$ in $H^3(\lambda)$ given by (\ref{2.7}) takes the form 
\be
F=\frac{1}{\lambda}\left(\ba{cc}
\bar{\alpha}_1 & \bar{\beta}_1 \\
\bar{\alpha}_2 & \bar{\beta}_2
\ea\right)\left(\ba{cc}
\vert\alpha_1\vert^2+\vert\beta_1\vert^2 & \bar{\alpha}_1\alpha_2+\bar{\beta}_1\beta_2 \\
\alpha_1\bar{\alpha}_2+\beta_1\bar{\beta}_2 & \vert\alpha_2\vert^2+\vert\beta_2\vert^2
\ea\right)\in\mathfrak{sl}(2,\mathbb{C})\label{4.a}
\ee
Hence, in the Lorentz space the immersion function $F$ expressed in terms of the $2\times2$ Hermitian matrices is given by (\ref{2.5}), where the components of the matrix $X$ are given by
\be
\ba{l}
X_0=\frac{1}{2\lambda}\left[\vert\alpha_1\vert^2+\vert\alpha_2\vert^2+\vert\beta_1\vert^2+\vert\beta_2\vert^2\right],\\
X_1=\frac{1}{\lambda}\mbox{Re}\left[\alpha_1\bar{\alpha}_2+\beta_1\bar{\beta}_2\right],\\
X_2=\frac{1}{\lambda}\mbox{Im}\left[\alpha_1\bar{\alpha}_2+\beta_1\bar{\beta}_2\right],\\
X_3=\frac{1}{2\lambda}\left[\vert\alpha_1\vert^2-\vert\alpha_2\vert^2+\vert\beta_1\vert^2-\vert\beta_2\vert^2\right].
\ea\label{4.b}
\ee
 with matrix elements given by (\ref{4.c}).

\section{Concluding remarks and future developments}
In this paper we study the Bryant representation \cite{Br,Br2} of conformal immersions of surfaces in the hyperbolic space $H^3(\lambda)\subset\mathbb{R}^{3,1}$. We describe the links between CMC-$\lambda$ surfaces and their corresponding minimal surfaces in Euclidean space $\mathbb{E}^3$, different from that found by Umehara and Yamada \cite{UY}. The most important advantage of our method is that we describe the Enneper-Weierstrass representation as a Sym-type immersion formula which completes the soliton-surface approach to minimal surfaces missing in earlier studies. Another important ingredient that we point out is the close connection between the Bryant representation of CMC-$\lambda$ surfaces with the theory of second-order linear ODEs. In particular we present such a surface related to the error function equation. Further investigation of the relation between various properties of ODEs and the Bryant representation of CMC-$\lambda$ surfaces will be performed in our future work.

\section*{Acknowledgements}
AD's work was supported in part by the Polish Ministry of Science and Higher Education grant No. N N202 174739. AMG's work was supported by a research grant from the Natural Sciences and Engineering Research Council of Canada.


\section*{References}
\begin{thebibliography}{10}
\expandafter\ifx\csname urlstyle\endcsname\relax
  \providecommand{\doi}[1]{doi:\discretionary{}{}{}#1}\else
  \providecommand{\doi}{doi:\discretionary{}{}{}\begingroup
  \urlstyle{rm}\Url}\fi
  
\bibitem{AL}
Ablowitz M., Ladik J.:  On solutions of a class of nonlinear partial difference equations. Stud. Appl. Math \textbf{57}, 1-12 (1977).

\bibitem{1}
Ablowitz M.J., Clarkson P. A.: Solitons. Nonlinear Evolution Equations and Inverse Scattering. Cambridge University Press, Cambridge (1991).

\bibitem{Abramowitz}
Abramowitz M., Stegun I. A.: Handbook of mathematical functions. National Bureau of Standards, Washington D.C (1964).

\bibitem{Bianchi}
Bianchi L.: Lezioni di Geometria Differentiale. 2nd ed., Spoerri, Pisa (1902).

\bibitem{3}
Bobenko A.I.: All constant mean curvature tori in $R^3$, $S^3$, $H^3$ in terms of theta-functions. Math. Ann. \textbf{290} 209-245 (1991).

\bibitem{Bob94}
Bobenko A.I.: Surfaces in terms of 2 by 2 matrices. Old and new integrable cases in Harmonic Maps and Integrable Systems. (Eds) Fordy A., Wood J. Braunschwieg, Vieweg (1994).

\bibitem{bo4}
Bobenko A.I., Seller R.: Discrete Integrable Geometry and Physics. Oxford Lect Ser in Math and Appl, Oxford (1999).

\bibitem{Bob}
Bobenko A.I., Eitner U.: Painlev\'e Equations in the Differential Geometry of Surfaces Lecture Notes in Mathematics. vol 1753, Springer, Berlin (2000).

\bibitem{Br}
Bryant R.L.: Surfaces of mean curvature one in hyperbolic space. Ast\'erisque 154-155, 321-347 (1987).

\bibitem{Br2}
Bryant R.L.: Surfaces in conformal geometry, in The mathematical heritage of Hermann Weyl. Proc. Sympos. Pure Math. \textbf{48}, Amer. Math. Soc. Providence 227-240 (1988).

\bibitem{4}
Cie\'sli\'nski J.: A generalized formula for integrable classes of surfaces in Lie algebras. J. Phys. Math. \textbf{38} 4255-4272 (1997).

\bibitem{5}
Cie\'sli\'nski J.: The spectral interpretation of $n$-spaces of constant negative curvature immersed in $\mathbb{R}^{2n-1}$. Phys. Lett. A \textbf{236} 425-430 (2007).

\bibitem{6}
Cie\'sli\'nski J.: A geometric interpretation of the spectral parameter for surfaces of constant mean curvature. J. Nonlin. Math. Phys. \textbf{13} 507-515 (2006).

\bibitem{7}
Cie\'sli\'nski J., Goldstein P., Sym A.: Isothermic surfaces in $E^3$ as soliton surfaces. Phys. Lett. A \textbf{205} 37-43 (1995).

\bibitem{9}
Doliwa A., Santini P.M.: An elementary geometric characterization of the integrable motions of a curve. Phys. Lett. A \textbf{185} 373-384 (1994).

\bibitem{10}
Doliwa A., Santini P.M.: Geometry of Discrete Curves and Lattices and Integrable Difference Equations, Discrete Integrable Geometry and Physics. Bobenko A and Seiler R (eds.), pp.139-154, Oxford University Press (1999).

\bibitem{DS92}
Doliwa., Sym A.: Constant mean curvature helicoids in $\mathbb{E}^3$ as an example of soliton surfaces, in Nonlinear Evolution Equations and Dynamical Systems (Eds) Boiti M., Martina L., Pempinelli F., World Scientific, Singapore (1992).

\bibitem{E}
Enneper E.: A Nachr. K\"onigl. Gesell. Wissensch Georg-Augusts-Univ. G\"ottingen \textbf{12}, 158, 421 (1868).

\bibitem{11}
Faddeev L.D., Takhtajan V.E.: Hamiltonian Methods in the Theory of Solitons. Springer, Berlin (1986).

\bibitem{12}
Fokas A.S., Gel'fand I.M.: Surfaces on Lie groups, on Lie algebras and their integrability. Commun. Math. Phys. \textbf{177} 203220 (1996).

\bibitem{13}
Fokas A.S., Gel'fand I.M., Finkel F., Liu Q.M.: A formula for constructing infinitely many surfaces on Lie algebras and integrable equations. Sel. Math. \textbf{6} 347-375 (2000).

\bibitem{GG}
Goldstein P., Grundland A.M.: Invariant recurrence relations for $\mathbb{C}P^{N-1}$ models. J. Phys. A: Math. Theor. \textbf{43} 265206, pp. 1-25 (2010).

\bibitem{14}
Grundland A.M., Post S.: Soliton surfaces associated with generalized symmetries of integrable equations. J. Phys A: Math. Theor.\textbf{44} 165203 31pp (2011).

\bibitem{K}
Kenmotsu K.: Weierstrass formula for surfaces of prescribed mean curvature. Math. Ann. 245, 89-99 (1979).

\bibitem{Kono}
Konopelchenko B.G., Landolfi G.: Induced surfaces and their integrable dynamics. Generalized Weierstrass representations in 4-D spaces and deformations via DS hierarchy. Studies Appl. Math. 104, 129-169 (1999).

\bibitem{Maurin}
Maurin K.: Analysis. D. Reidel Publ. Comp., Dordrecht (1976).

\bibitem{15}
Nitsche J.C.C.: Vorlesungen \"uber Minimalfl\"achen. Grundlehren 199, Springer, Berlin (1975).

\bibitem{16}
Osserman R.: A survey of minimal surfaces. Van Nostrand Reinhold Co., New York (1969).

\bibitem{Spirak}
Spivak M.: A Comprehensive Introduction to Differential Geometry. Publish or Perish Inc., Houston (1979).

\bibitem{17}
Sym A.: Soliton theory is surface theory. Institute of Theoretical Physics, Warsaw University, preprint IFT/11/81 Warsaw (1982).

\bibitem{18}
Sym A.: Soliton surfaces. Lett. Nuovo Cimento \textbf{33} 394-400  (1982).

\bibitem{19}
Sym A.: Soliton Surfaces and Their Applications. Lecture Notes in Physics, Vol. 239, p.154, Springer, Berlin (1985).

\bibitem{UY} Umehara M., Yamada K.: A parametrization of the Weierstrass formulae and perturbation of complete minimal surfaces in $\mathbb{R}^3$ into the hyperbolic 3-space. J. Reine Angew Math. 432, 93-116 (1992).

\bibitem{W}
Weierstrass K.: Fortsetzung der Untersuchung uber die Minimalflchen. in Mathematische Werke. Vol 3, Verlagsbchhandlung, Milleshein, 219 (1866).

\bibitem{Yang}
Yang K.: Complete and Compact Minimal Surfaces. Vol 54, Kluwer Acad. Publ., Dordrecht (1989).

\bibitem{ZS}
Zakharov V.E., Shabat A.B.: Integration of nonlinear equations of mathematical physics by the method of inverse scattering I and II. Func. Anal. Appl. 8, 226-235 (1974); 13, 13-22 (1978).

\end{thebibliography}
\end{document}